\documentclass[onecolumn, conference]{IEEEtran}
\usepackage[noadjust]{cite}
\usepackage[T1]{fontenc}
\usepackage{lmodern}
\usepackage{amssymb,amsmath}
\usepackage{amsthm}
\usepackage{graphicx}
\usepackage{caption}
\usepackage{subcaption}
\usepackage{relsize}

\usepackage[shortlabels]{enumitem}

\usepackage[table]{xcolor}
\usepackage{array}[2003/12/17]
\usepackage{calc}

\newlength\celldim \newlength\fontheight \newlength\extraheight
\newcounter{sqcolumns}

\newcolumntype{S}{ @{}
  >{\centering \rule[-0.5\extraheight]{0pt}{\fontheight + \extraheight}}
  p{\celldim} @{} }

\newcolumntype{Z}{ @{} >{\centering} p{\celldim} @{} }

\newenvironment{squarecells}[1]
  {
   \setlength{\tabcolsep}{0em} 
   \setlength\celldim{1.5em}%
   \settoheight\fontheight{A}%
   \setlength\extraheight{\celldim - \fontheight}%
   \setcounter{sqcolumns}{#1 - 1}%
   \begin{tabular}{S*{\value{sqcolumns}}{Z}}}
  {\end{tabular}}

\newcommand\nl{\tabularnewline}

\newcommand{\x}{\times}

\renewcommand{\bar}{\overline}
\renewcommand{\epsilon}{\varepsilon}

\newcommand{\set}[1]{\{#1\}}

\usepackage{xcolor}

\renewcommand{\proof}{{\bf Proof. }}

\providecommand{\qed}{$\square$}

\newtheorem{lemma}{Lemma}
\newtheorem{theorem}{Theorem}

\newtheoremstyle{remarkstyle} 
    {\topsep}                    
    {\topsep}                    
    {}                   
    {}                           
    {\itshape}                   
    {.}                          
    {.5em}                       
    {}  
\theoremstyle{remarkstyle}
\newtheorem{remark}{Remark}

\newcommand{\beq}{\begin{equation}}
\newcommand{\eeq}{\end{equation}}
\newcommand{\supth}{^{th}}

\let\Oldforall\forall
\renewcommand{\forall}{~\Oldforall} 

\let\Oldinf\inf
\renewcommand{\inf}{\Oldinf\limits}
\let\Oldsup\sup
\renewcommand{\sup}{\Oldsup\limits}

\IfFileExists{microtype.sty}{\usepackage{microtype}}{}
\usepackage{hyperref}
\title{Optimal Systematic Distributed Storage Codes with Fast Encoding}

\usepackage{physics}
\usepackage{blkarray}
\usepackage[normalem]{ulem}
\usepackage{mathtools}

\date{Aug 10, 2015}

\author{
\IEEEauthorblockN{{Preetum Nakkiran, KV Rashmi, Kannan Ramchandran}}
\IEEEauthorblockA{{\it University of California, Berkeley}}
}

\setlist[enumerate]{label=(\arabic*)}
\begin{document}

\newcommand{\C}{\mathcal{C}}
\newcommand{\incl}[1][]{\xhookrightarrow{#1}}
\newcommand{\too}[1][]{\xrightarrow{#1}}
\newcommand{\fe}{f_{e}}
\newcommand{\fremap}{f_{S}}
\newcommand{\finc}{f_{\iota}}
\newcommand{\bM}{\bar{M}}
\newcommand{\Sa}{{S^a}}
\newcommand{\Sb}{{S^b}}
\newcommand{\bSa}{\bar{S^a}}
\newcommand{\bSb}{\bar{S^b}}
\newcommand{\Ck}{{C_k}}

\maketitle

\newcommand{\pmmsr}{PM}

\begin{abstract}
Erasure codes are being increasingly used in distributed-storage systems
in place of data-replication, since they provide the same level of reliability with
much lower storage overhead.
We consider the problem of constructing explicit erasure codes for distributed storage
with the following desirable properties motivated by practice:
(i) \emph{Maximum-Distance-Separable (MDS)}:
to provide maximal reliability at minimum storage overhead,
(ii) \emph{Optimal repair-bandwidth}:
to minimize the amount of data needed to be transferred to repair a failed node
from remaining ones,
(iii) \emph{Flexibility in repair}:
to allow maximal flexibility in selecting subset of nodes to use for repair,
which includes not requiring that all surviving nodes be used for repair,
(iv) \emph{Systematic Form}:
to ensure that the original data exists in uncoded form, and
(v) \emph{Fast encoding}: to minimize the cost of generating
encoded data (enabled by a \emph{sparse} generator matrix).
Existing constructions in the literature satisfy only strict subsets of these desired
properties.

This paper presents the first explicit code construction
which theoretically guarantees all the
five desired properties \emph{simultaneously}.
Our construction builds on a powerful class of codes called \emph{Product-Matrix (PM) codes}.
PM codes satisfy properties
(i)-(iii), and either (iv) or (v),  but not both simultaneously.
Indeed, native PM codes have inherent structure that leads to sparsity,
but this structure is destroyed when the codes are made systematic.
We first present an analytical framework for understanding the interaction
between the design of PM codes and the systematic property.
Using this framework, we provide an explicit code construction
that simultaneously achieves all the above desired properties.
We also present general ways of transforming existing
storage and repair optimal codes
to enable fast encoding through sparsity.
In practice, such sparse codes result in encoding speedup
by a factor of about 4 for typical parameters.
\end{abstract}

\section{Introduction}
Erasure codes are being increasingly used in distributed-storage systems
instead of replication, since they provide the same level of reliability with
much less storage overhead.
Large scale distributed-storage systems have many practical requirements that
guide the design of distributed-storage codes.

\begin{figure}[h]
\centering
\hfill
\begin{subfigure}[h]{0.4\textwidth}
\centering
\includegraphics[height=2in]{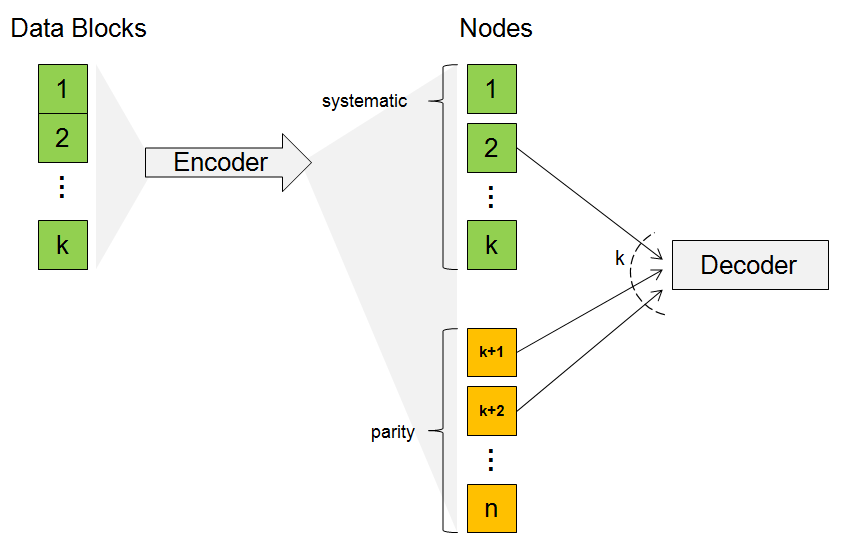}
\caption{}
\label{fig:code}
\end{subfigure}
\hfill
\begin{subfigure}[h]{0.4\textwidth}
\centering
\includegraphics[height=2in]{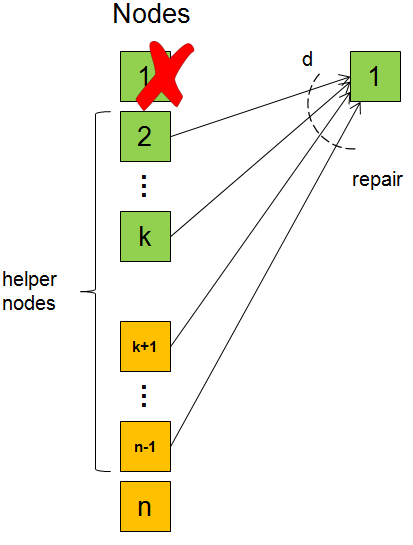}
\caption{}
\label{fig:repair}
\hfill
\end{subfigure}
\caption{(a) Encoding and decoding for an $[n, k]$ systematic MDS code, (b) Node repair: Connecting to $d=(n-2)$ helper nodes
to repair failed node 1.}
\end{figure}

In large-scale systems, storage is a critical resource.
For this reason, \emph{Maximum-Distance-Separable (MDS)} codes such as
Reed-Solomon codes, which require the minimal storage overhead to achieve a
desired level of reliability, are a popular choice
\cite{gfs, hdfs-raid1, hdfs-raid2}.
An $[n, k]$ MDS code allows the data to be stored across $n$ nodes
such that
the entire data can be recovered from the encoded data stored
in any $k$ (out of $n$) nodes.
This is depicted in Figure~\ref{fig:code}.
Another critical resource in distributed-storage systems is network bandwidth.
In large-scale systems, failures are the norm rather than the exception,
and repair operations run continuously in the background~\cite{rashmi2013hotstorage}.
When nodes fail, they must be repaired by
downloading some data from the remaining nodes. These nodes are termed
\textit{helper nodes}. Figure~\ref{fig:repair} depicts a repair operation where
node $1$ is being repaired with the help of nodes $\{2,\ldots,n-1\}$. In
large-scale systems, the repair operations consume a significant amount of
network bandwidth, and this has been one of main
deterrents to using classical MDS erasure codes in such systems~\cite{rashmi2013hotstorage}.
Hence, it is important for storage codes to also minimize the amount of bandwidth
consumed during repair.

Another important system consideration
is that the code not force the requirement that all surviving $(n-1)$
nodes be needed to repair a single failed node.
If $d$ denotes the number of helper nodes required for repair,
then this property requires $d < (n-1)$, as illustrated in
Figure~\ref{fig:repair}.
This property is crucial to allow redundant requests
to be sent during a repair operation, which is an effective approach to reducing
latency in practical systems~\cite{vulimiri2012more, liang2014fast,ananthanarayanan2012let,dean2013tail,shah2013redundant}.
That is, a failed node can request help from many helpers, and can repair as
soon as enough nodes respond.
This property is even more critical for \emph{degraded reads}~\cite{rashmi2013hotstorage},
where a repair operation is performed to serve a read request for data stored in a busy or otherwise unavailable node. Latency is crucial for degraded reads to meet the service
level agreement in large scale systems.

Another practical requirement of storage codes is that of being in \emph{systematic} form.
That is, the original data must exist in the system in uncoded form.
Figure~\ref{fig:code} shows a systematic code wherein the first $k$ nodes store the original data.
This is essential when serving read requests, since
if the code is systematic, read requests can be served by simply reading the
data in systematic nodes.
Otherwise the system must perform a decoding operation to retrieve
the original data for every read request. 

Finally, one of the most frequent operations performed in many distributed-storage systems
is the encoding of new data entering the system.
This encoding cost is a non-issue when using replication,
but can be significant when using erasure codes.
Thus it is desirable for the code to support fast encoding operations.
For linear codes, encoding the original data can be represented as multiplication
between a generator matrix and the data vector \cite{lin-costello}.
This encoding operation will be fast if the generator matrix is sparse,
since this reduces the number of computations performed.
Informally, the sparsity of the generator matrix dictates how many data symbols
need to be touched in order to generate each encoded symbol.

This forms the motivation for this paper:
to construct storage codes that satisfy \emph{all} the above
system-driven constraints.
That is, storage codes having the following five properties:
(i) Minimum storage for a targeted level of reliability (MDS),
(ii) Minimal repair bandwidth,
(iii) Flexible repair parameters: $d < (n-1)$,
(iv) Systematic form of encoded data, and
(v) Fast encoding, enabled by a sparse generator matrix.


There has been considerable interest in the recent past in constructing such erasure codes
for distributed storage
\cite{MBR_pt, MISERjournal, tamo2013zigzag,
    papailiopoulos2013repairTransactions,
cadambe2011polynomial}.
However, to the best of our knowledge,
all existing constructions in the literature address only a
\emph{strict subset} of the above desired properties.
This paper presents the first explicit codes which theoretically guarantee all the
five desired properties simultaneously.

Our constructions are based on a powerful class of storage codes
called \emph{Product-Matrix (PM)} codes \cite{PMC}.
PM codes are MDS
\footnote{
    We use ``PM codes'' here to refer to the MDS version of
    Product-Matrix codes, termed PM-MSR in \cite{PMC}.
}
, and hence are optimal w.r.t. storage overhead.
They also have optimal bandwidth consumed during repair,
since they meet the lower-bound presented in \cite{netcoding}.
PM codes belong to a general class of codes known as
\emph{Regenerating codes} \cite{netcoding}, which meet this lower-bound.
Moreover, PM codes support a wide range of values for
$d: (2k-2) \leq d \leq (n-1)$.
Finally, the special structure of PM codes makes their generator matrix sparse,
leading to fast encoding \cite{jiekak2012system}.
Thus PM codes satisfy Properties (i)-(iii) and (v). 

Native PM codes, however, are not systematic.
They can be converted to systematic form using
a generic transformation termed ``systematic-remapping'' \cite{PMC}.
However, this remapping does not respect the inherent structure of PM codes,
and \emph{thus often destroys its sparsity.}
Thus naively performing a remapping transform causes PM codes
to be systematic, at the expense of fast encoding.
For example, an $[n,k,d=2k-1]$ PM code requires a
block-length of $k^2$ symbols.
That is, each stored symbol can be, in general, a function of
up to $k^2$ data symbols.
However, due to the sparse structure of native PM codes, each stored symbol
is a function of only $O(k)$ of these data symbols.
This is no longer true after systematic remapping, and in general each parity
symbol becomes a (dense) function of $k^2$ symbols.
This results in significantly higher encoding time for systematic PM codes
constructed in this manner \cite{jiekak2012system, fast-pm-codes}.

In this paper, first, we present an analytical framework for studying and understanding
the interaction between the design of PM codes and the systematic-remapping transformation.
Using this, we provide an explicit construction of PM codes which remains
sparse after systematic-remapping, for $d=(2k-2)$.
In particular, each parity symbol in this construction depends on only $d=O(k)$ data symbols.
\footnote{
    Note that for a systematic $[n,k,d]$ MDS code,
    a sparsity of at least $k$ symbols is necessary for each encoded symbol in the parity nodes.}

Second, we consider the sparsity of codes supporting
\emph{repair-by-transfer.}
A node assisting in a repair operation is said to perform repair-by-transfer
if it does not perform any computation, and merely
transfers one of its stored symbols to
the failed node \cite{allerton09}.
Storage codes which support repair-by-transfer are appealing in practice, since they
also minimize the amount of data read during repairs.
There have been a number of works in the recent past
on constructing such storage codes
\cite{allerton09, fast,
rouayheb2010fractional,
pawardress,
hu2013analysis,
rashmi2014hitchhiker}.
We show that a particular type of repair-by-transfer property leads to sparsity
in \emph{any} MDS regenerating code.
This provides a general way of constructing sparse MDS regenerating codes.

Third, using the above result,
we construct explicit sparse systematic PM codes
for all $d \geq (2k-2)$.
For example, the generator matrix of a $[n=17,k=8,d=15]$
systematic-remapped PM code as in \cite{PMC} is $\sim 11\%$ sparse,
while our construction is $\sim 77\%$ sparse.

We note that the construction provided in this paper
is similar to the codes considered in \cite{fast},
wherein the authors present codes supporting repair-by-transfer for
achieving savings in disk I/O.
For $d=(2k-2)$, the construction provided in the present paper
is also similar to the recent construction in \cite{fast-pm-codes}
by Le Scouarnec.
In \cite{fast-pm-codes}, the author presents a sparse \pmmsr{} code
and computationally validates its properties
for a fixed range of $k$.
In fact, the results presented in this paper
provide a theoretical proof of sparsity for
the constructions in both the above works
\cite{fast, fast-pm-codes}.

The remainder of this paper is organized as follows:
Section~\ref{sec:background} contains a review of
Product-Matrix codes, systematic-remapping,
and other necessary background and notation.
Section~\ref{sec:example} contains a motivating example. 
Section~\ref{sec:step1} illustrates the main ideas of our approach
to understanding sparsity, by showing that a simple form of \pmmsr{} encoding matrix leads to
partial sparsity. These techniques are extended in Section~\ref{sec:explicit1},
to give an explicit construction of sparse systematic \pmmsr{} codes for $d=(2k-2)$.
In Section~\ref{sec:general} we consider more general regenerating codes,
and show that regenerating codes possessing a certain
repair-by-transfer property are necessarily sparse.
We apply this in Section~\ref{sec:explicit2} to construct
explicit sparse systematic \pmmsr{} codes for $d \geq (2k-2)$.
Finally in Section~\ref{sec:equivalence}, we show that the two
presented constructions of sparse \pmmsr{} codes for $d=(2k-2)$ are in fact
equivalent in a certain sense.

\section{Background}
\label{sec:background}


\subsection{Product-Matrix Codes}
\label{sec:PMC}
Product-Matrix (\pmmsr{}) codes \cite{PMC}
are an explicit family of linear MDS codes which minimize bandwidth consumed
in repair, and exist for all $[n,k,d \geq 2k-2]$.

Let the message to be stored consist of $B$ symbols from the finite field
$\mathbb{F}_q$.
An $[n, k, d](\alpha)$
\pmmsr{} code allows the message
to be stored across $n$ nodes, each storing $\alpha$ encoded symbols.
All the $B$ symbols can be recovered from the data stored in any $k$ of the
total $n$ nodes.
Further, any node's data may be exactly recovered by connecting to
any $d$ other nodes, and downloading one symbol from each.
These $d$ nodes are known as ``helper nodes.''
The symbols transferred from a helper node during node repair
will be a linear function of the data stored in it.
\pmmsr{} codes are storage-optimal and hence
\beq
B = k \alpha.
\eeq
The parameter $\alpha$ is induced by $[n,k,d]$ as
\beq
\alpha = d-k+1.
\eeq

We now describe the construction of \pmmsr{} codes.
In general, a \pmmsr{} code is described by
an $(n \times d)$ \emph{encoding matrix} $\Psi$ and a $(d \times \alpha)$
\emph{message matrix} $M$,
yielding an $(n \times \alpha)$
\emph{code matrix} $C$ defined by
\beq
\label{eqn:CPsiM}
C:=\Psi M.
\eeq
Let $c_i^T$ denote the $i^{th}$ row of the code matrix $C$.
Then the $i^{th}$ node stores $c_i^T = \psi_i^T M$.

Here we review \pmmsr{} codes for $d=(2k-2)$,
but the construction can applied to $d > (2k-2)$
by the \emph{shortening} procedure of \cite{PMC},
which we review in Section~\ref{sec:shorten}.
\footnote{
Constructions without puncturing were subsequently shown in \cite{unified}
and \cite{rskgeneral}.}

For $d=(2k-2)$, we have
$\alpha = (d-k+1) = (k-1)$.
For these parameters, the encoding matrix $\Psi$  is of the form:
\beq
\label{eqn:psi}
\Psi = \mqty[\Phi & \Lambda \Phi]
\eeq
where $\Phi$ is an $(n \times \alpha)$ matrix and $\Lambda$ is an $(n \times n)$
diagonal matrix, with the following properties:
\begin{enumerate}
    \item Any $\alpha$ rows of $\Phi$ are linearly independent
    \item Any $d$ rows of $\Psi$ are linearly independent
    \item The diagonal elements of $\Lambda$ are all distinct.
\end{enumerate}

These requirements can be met, for example, by choosing $\Psi$ to be a Vandermonde matrix
with elements chosen carefully to satisfy the third condition.

We will now specify the structure of the message matrix $M$.
Recall for $d=(2k-2)$, we have $\alpha=(k-1)$, $d= 2\alpha$, and $B = k\alpha = \alpha(\alpha+1)$.
The $(d \times \alpha)$ message matrix $M$ is constructed as
\beq
M = \mqty[\Sa \\ \Sb]
\eeq
where $\Sa$ and $\Sb$ are $(\alpha \times \alpha)$ symmetric matrices.
The matrices $\Sa$ and $\Sb$ together have precisely $\alpha(\alpha+1)$ distinct entries,
which are now populated by the $B=\alpha (\alpha+1)$ message symbols.

Let $\psi_i^T$ denote the $i^{th}$ row of $\Psi$,
and $\phi_i^T$ denote the $i^{th}$ row of $\Phi$.
Thus, under this encoding mechanism, node $i~(1 \leq i \leq n)$, stores the $\alpha$ symbols
\beq
\label{eqn:node-stores}
c_i^T = {\psi}_i^T M = {\phi}_i^T \Sa+ \lambda_{i} {\phi}_i^T \Sb~.
\eeq

Under this encoding, the data in any $k$ nodes
suffice to reconstruct the $B=k\alpha$ message symbols.
The original paper \cite{PMC} presents an explicit reconstruction algorithm
for general \pmmsr{} codes, relying on Properties 1 and 3 above.

\pmmsr{} codes allow repair of any failed node,
by downloading
one symbol from any $d$ other helper nodes.
For repairing node $f$, helper node $i$ sends the single symbol
\beq
\label{eqn:repair}
c_i^T \phi_f = \psi_i^T M \phi_f.
\eeq
Upon receiving $d$ such helper symbols, failed node $f$ will have
$\Psi_d M \phi_f$, where $\Psi_d$ is some $d$ rows of $\Psi$.
It can then invert $\Psi_d$ (by Property 2) to compute
\beq
M\phi_f = \mqty[\Sa \phi_f \\ \Sb \phi_f]~.
\eeq
And thus can recover its data as
\beq
c_f^T
= (\Sa \phi_f)^T + \lambda_f (\Sb \phi_f)^T
= {\phi}_f^T \Sa+ \lambda_{f} {\phi}_f^T \Sb
\eeq
(follows by symmetry of the matrices $\Sa$ and $\Sb$).

\subsection{Systematic Codes and Remapping}
\label{sec:sys-remap}
It is often desirable to have
the $B$ original message symbols
included in the encoded symbols
(in uncoded form).
Such codes are called \emph{systematic codes}.
Throughout the paper we consider systematic codes in which
the first $k$ nodes store the uncoded symbols.
These nodes are thus referred to as ``systematic nodes.''

Any linear MDS erasure code can be generically transformed into a systematic code,
as follows.
First, any linear code taking $B$ message symbols to $n\alpha$ encoded
symbols can be represented by an $(n\alpha \x B)$ \emph{generator matrix} $G$,
such that for a message-vector $m$ of length $B$, the encoded $n\alpha$ symbols are
given by $Gm$.

A code can be made systematic
through a ``systematic remapping'':
Let $G_k$ be a $(B \x B)$ matrix consisting of the
first $B$ rows of the original generator matrix $G$.
To encode message $m$, first ``remap'' the message vector to
$\bar m := G_k^{-1}m$, then encode as $G \bar m$.
Consider the resulting first $B$ encoded symbols: 
the message $m$ is first transformed by $G_k^{-1}$,
then transformed by $G_k$ during encoding.
Therefore the first $B$ encoded symbols are exactly the message symbols $m$,
making the code systematic.
Notice that the entire encoding operation now is equivalent to encoding
the original message $m$ with generator matrix $G_{sys} := G G_k^{-1}$,
which will have the first $(B \x B)$ block as identity by construction.

Observe that the systematic remapping operation applies $G_k^{-1}$,
and hence can be thought of as decoding the message from the first $k$ nodes
under the original encoding with generator matrix $G$.

The above transform can be applied to the vanilla \pmmsr{} codes
discussed in Section~\ref{sec:PMC} and \cite{PMC},
to yield systematic \pmmsr{} codes.
However, as shown in the examples below, applying systematic remapping
to traditional \pmmsr{} codes often destroys their sparsity -- leading to
increased computational complexity.

\subsection{Notation}
We will use the concept of an \emph{inclusion map}.
In general, an inclusion map is a map which injectively embeds one space into
another space, by
simply changing representation (not performing any non-trivial transformation).
For example, the following is an inclusion map from vectors of length 3 to symmetric
$(2 \x 2)$ matrices:
$$\mqty[a \\ b \\ c] \incl \mqty[a & b \\ b & c]$$

Inclusion maps will be denoted by hooked arrows ($\incl$) as above.

For notational simplicity, we will often abuse notation by using the same symbols
to denote a space as well as a vector in the space.
For example, the systematic-remapping transformation of a message vector $m$,
as in Section~\ref{sec:sys-remap}, will be written as a function
$f: m \to \bar m$.

The $(i,j)^{th}$ entry of a matrix $M$
is denoted $M_{i,j}$.
All vectors are column-vectors unless otherwise noted, and $^T$ denotes
transpose throughout.

\section{Motivating Example}
\label{sec:example}

\subsection{Example}
\label{sec:example_ex}
To better understand the issues of sparsity and systematic remapping in \pmmsr{}
codes, let us consider a particular $[n=8, k=4, d=6]$ \pmmsr{} code.
For these parameters, each node stores $\alpha = 3$ symbols, and the number of message symbols
is $B = 12$.
Let $\set{m_0, \dots, m_{11}}$ denote these message symbols.
Let us work in field $\mathbb{F}_{11}$
\footnote{This is the smallest prime field which will allow the \pmmsr{} construction of
\cite{PMC} for this parameter regime.}.
As described in Section~\ref{sec:PMC}, we have:

\setcounter{MaxMatrixCols}{20}

$$
\Psi=
\mqty[
1 & 1 & 1 & 1 & 1 & 1\\
2 & 4 & 8 & 5 & 10 & 9\\
3 & 9 & 5 & 4 & 1 & 3\\
4 & 5 & 9 & 3 & 1 & 4\\
5 & 3 & 4 & 9 & 1 & 5\\
6 & 3 & 7 & 9 & 10 & 5\\
7 & 5 & 2 & 3 & 10 & 4\\
8 & 9 & 6 & 4 & 10 & 3\\
]
, \qquad\qquad
M =
\mqty[  m_0 & m_1 & m_2\\
        m_1 & m_3 & m_4\\
        m_2 & m_4 & m_5\\
    \hline
        m_6 & m_7 &  m_8\\
        m_7 & m_9 &  m_{10}\\
        m_8 & m_{10} & m_{11}]
$$
Recall from Section~\ref{sec:PMC} that node $i$ stores the $i$-th row of $\Psi$ times $M$,
so the entire code is $C = \Psi M$.

As in Section~\ref{sec:sys-remap}, we can generically represent the encoding operation
as an $(n\alpha \x B) = (24 \x 12)$ generator matrix $G$ times the message \emph{vector} $m$,
with entries $m_i$.
That is, we can ``unwrap'' the matrix-matrix multiplication $C = \Psi M$
into each of $n\alpha = 24$ encoded symbols.
For example, the first $\alpha=3$ rows of $G$ correspond to the 3 linear
combinations stored by the first node:
$$
\mqty[
1 & 1 & 1 & 0 & 0 & 0 & 1 & 1 & 1 & 0 & 0 & 0\\
0 & 1 & 0 & 1 & 1 & 0 & 0 & 1 & 0 & 1 & 1 & 0\\
0 & 0 & 1 & 0 & 1 & 1 & 0 & 0 & 1 & 0 & 1 & 1\\
]
$$
And the next $3$ rows of $G$ correspond to the 3 linear combinations
stored by the second node:
$$
\mqty[
2 & 4 & 8 & 0 & 0 & 0 & 5 & 10 & 9 & 0 & 0 & 0\\
0 & 2 & 0 & 4 & 8 & 0 & 0 & 5 & 0 & 10 & 9 & 0\\
0 & 0 & 2 & 0 & 4 & 8 & 0 & 0 & 5 & 0 & 10 & 9\\
]
$$

Notice that the submatrix of the generator matrix corresponding to each node
is $d$-sparse,
with the same sparsity pattern.
The entire generator matrix and its sparsity pattern are as follows:

\beq
\label{eqn:Gorig}
G=
\smqty[
1 & 1 & 1 & 0 & 0 & 0 & 1 & 1 & 1 & 0 & 0 & 0\\
0 & 1 & 0 & 1 & 1 & 0 & 0 & 1 & 0 & 1 & 1 & 0\\
0 & 0 & 1 & 0 & 1 & 1 & 0 & 0 & 1 & 0 & 1 & 1\\
2 & 4 & 8 & 0 & 0 & 0 & 5 & 10 & 9 & 0 & 0 & 0\\
0 & 2 & 0 & 4 & 8 & 0 & 0 & 5 & 0 & 10 & 9 & 0\\
0 & 0 & 2 & 0 & 4 & 8 & 0 & 0 & 5 & 0 & 10 & 9\\
3 & 9 & 5 & 0 & 0 & 0 & 4 & 1 & 3 & 0 & 0 & 0\\
0 & 3 & 0 & 9 & 5 & 0 & 0 & 4 & 0 & 1 & 3 & 0\\
0 & 0 & 3 & 0 & 9 & 5 & 0 & 0 & 4 & 0 & 1 & 3\\
4 & 5 & 9 & 0 & 0 & 0 & 3 & 1 & 4 & 0 & 0 & 0\\
0 & 4 & 0 & 5 & 9 & 0 & 0 & 3 & 0 & 1 & 4 & 0\\
0 & 0 & 4 & 0 & 5 & 9 & 0 & 0 & 3 & 0 & 1 & 4\\
    \hline
5 & 3 & 4 & 0 & 0 & 0 & 9 & 1 & 5 & 0 & 0 & 0\\
0 & 5 & 0 & 3 & 4 & 0 & 0 & 9 & 0 & 1 & 5 & 0\\
0 & 0 & 5 & 0 & 3 & 4 & 0 & 0 & 9 & 0 & 1 & 5\\
6 & 3 & 7 & 0 & 0 & 0 & 9 & 10 & 5 & 0 & 0 & 0\\
0 & 6 & 0 & 3 & 7 & 0 & 0 & 9 & 0 & 10 & 5 & 0\\
0 & 0 & 6 & 0 & 3 & 7 & 0 & 0 & 9 & 0 & 10 & 5\\
7 & 5 & 2 & 0 & 0 & 0 & 3 & 10 & 4 & 0 & 0 & 0\\
0 & 7 & 0 & 5 & 2 & 0 & 0 & 3 & 0 & 10 & 4 & 0\\
0 & 0 & 7 & 0 & 5 & 2 & 0 & 0 & 3 & 0 & 10 & 4\\
8 & 9 & 6 & 0 & 0 & 0 & 4 & 10 & 3 & 0 & 0 & 0\\
0 & 8 & 0 & 9 & 6 & 0 & 0 & 4 & 0 & 10 & 3 & 0\\
0 & 0 & 8 & 0 & 9 & 6 & 0 & 0 & 4 & 0 & 10 & 3\\
]
\qquad\sim\qquad
\smqty[
* & * & * &   &   &   & * & * & * &   &   &  \\
  & * &   & * & * &   &   & * &   & * & * &  \\
  &   & * &   & * & * &   &   & * &   & * & *\\
* & * & * &   &   &   & * & * & * &   &   &  \\
  & * &   & * & * &   &   & * &   & * & * &  \\
  &   & * &   & * & * &   &   & * &   & * & *\\
* & * & * &   &   &   & * & * & * &   &   &  \\
  & * &   & * & * &   &   & * &   & * & * &  \\
  &   & * &   & * & * &   &   & * &   & * & *\\
* & * & * &   &   &   & * & * & * &   &   &  \\
  & * &   & * & * &   &   & * &   & * & * &  \\
  &   & * &   & * & * &   &   & * &   & * & *\\
    \hline
* & * & * &   &   &   & * & * & * &   &   &  \\
  & * &   & * & * &   &   & * &   & * & * &  \\
  &   & * &   & * & * &   &   & * &   & * & *\\
* & * & * &   &   &   & * & * & * &   &   &  \\
  & * &   & * & * &   &   & * &   & * & * &  \\
  &   & * &   & * & * &   &   & * &   & * & *\\
* & * & * &   &   &   & * & * & * &   &   &  \\
  & * &   & * & * &   &   & * &   & * & * &  \\
  &   & * &   & * & * &   &   & * &   & * & *\\
* & * & * &   &   &   & * & * & * &   &   &  \\
  & * &   & * & * &   &   & * &   & * & * &  \\
  &   & * &   & * & * &   &   & * &   & * & *\\
]
\eeq

This code is not systematic, since it does not
contain the uncoded message symbols.
To make it systematic, we perform the systematic remapping
of Section~\ref{sec:sys-remap}:
Let $G_k$ be the $(B \x B)$ matrix consisting of the first $B$ rows of the
generator matrix $G$
(above the line in \eqref{eqn:Gorig}).
The systematic generator matrix is
$G_{sys} = G G_k^{-1}$, which in our case is:

\beq
\label{eqn:Gsys}
G_{sys}=
\smqty[
1 & 0 & 0 & 0 & 0 & 0 & 0 & 0 & 0 & 0 & 0 & 0\\
0 & 1 & 0 & 0 & 0 & 0 & 0 & 0 & 0 & 0 & 0 & 0\\
0 & 0 & 1 & 0 & 0 & 0 & 0 & 0 & 0 & 0 & 0 & 0\\
0 & 0 & 0 & 1 & 0 & 0 & 0 & 0 & 0 & 0 & 0 & 0\\
0 & 0 & 0 & 0 & 1 & 0 & 0 & 0 & 0 & 0 & 0 & 0\\
0 & 0 & 0 & 0 & 0 & 1 & 0 & 0 & 0 & 0 & 0 & 0\\
0 & 0 & 0 & 0 & 0 & 0 & 1 & 0 & 0 & 0 & 0 & 0\\
0 & 0 & 0 & 0 & 0 & 0 & 0 & 1 & 0 & 0 & 0 & 0\\
0 & 0 & 0 & 0 & 0 & 0 & 0 & 0 & 1 & 0 & 0 & 0\\
0 & 0 & 0 & 0 & 0 & 0 & 0 & 0 & 0 & 1 & 0 & 0\\
0 & 0 & 0 & 0 & 0 & 0 & 0 & 0 & 0 & 0 & 1 & 0\\
0 & 0 & 0 & 0 & 0 & 0 & 0 & 0 & 0 & 0 & 0 & 1\\
4 & 2 & 1 & 2 & 7 & 8 & 0 & 0 & 3 & 2 & 5 & 7\\
8 & 0 & 0 & 7 & 1 & 9 & 10 & 0 & 3 & 9 & 2 & 5\\
1 & 6 & 10 & 7 & 8 & 10 & 0 & 10 & 4 & 10 & 3 & 9\\
5 & 7 & 4 & 2 & 3 & 0 & 1 & 3 & 5 & 4 & 4 & 9\\
9 & 2 & 10 & 9 & 0 & 3 & 9 & 4 & 8 & 3 & 4 & 4\\
9 & 2 & 9 & 3 & 0 & 0 & 10 & 2 & 7 & 8 & 7 & 2\\
10 & 7 & 7 & 4 & 6 & 8 & 5 & 10 & 5 & 10 & 0 & 4\\
5 & 7 & 4 & 4 & 0 & 8 & 7 & 4 & 4 & 8 & 10 & 0\\
9 & 9 & 0 & 6 & 8 & 9 & 4 & 2 & 7 & 0 & 8 & 3\\
7 & 5 & 0 & 5 & 4 & 6 & 2 & 7 & 2 & 10 & 3 & 7\\
5 & 8 & 5 & 7 & 6 & 0 & 1 & 9 & 9 & 0 & 10 & 3\\
8 & 0 & 8 & 4 & 6 & 10 & 5 & 3 & 8 & 6 & 3 & 6\\
]
\qquad\sim\qquad
\smqty[
* &   &   &   &   &   &   &   &   &   &   &  \\
  & * &   &   &   &   &   &   &   &   &   &  \\
  &   & * &   &   &   &   &   &   &   &   &  \\
  &   &   & * &   &   &   &   &   &   &   &  \\
  &   &   &   & * &   &   &   &   &   &   &  \\
  &   &   &   &   & * &   &   &   &   &   &  \\
  &   &   &   &   &   & * &   &   &   &   &  \\
  &   &   &   &   &   &   & * &   &   &   &  \\
  &   &   &   &   &   &   &   & * &   &   &  \\
  &   &   &   &   &   &   &   &   & * &   &  \\
  &   &   &   &   &   &   &   &   &   & * &  \\
  &   &   &   &   &   &   &   &   &   &   & *\\
* & * & * & * & * & * &   &   & * & * & * & *\\
* &   &   & * & * & * & * &   & * & * & * & *\\
* & * & * & * & * & * &   & * & * & * & * & *\\
* & * & * & * & * &   & * & * & * & * & * & *\\
* & * & * & * &   & * & * & * & * & * & * & *\\
* & * & * & * &   &   & * & * & * & * & * & *\\
* & * & * & * & * & * & * & * & * & * &   & *\\
* & * & * & * &   & * & * & * & * & * & * &  \\
* & * &   & * & * & * & * & * & * &   & * & *\\
* & * &   & * & * & * & * & * & * & * & * & *\\
* & * & * & * & * &   & * & * & * &   & * & *\\
* &   & * & * & * & * & * & * & * & * & * & *\\
]
\eeq

Notably, the parity nodes are now almost entirely dense.
\footnote{In general, they will be entirely dense -- the small sparsities here
are incidental, due to small field size.}

\subsection{Discussion}
As seen here, traditional \pmmsr{} codes begin sparse, but become dense after
systematic-remapping.
We may expect this, since
\emph{the initial sparsity of \pmmsr{} codes comes from their
product-matrix structure, but the systematic-remapping operates generically
on linear codes, not necessarily respecting the product-matrix
structure.}
To address this, we need to understand the effect of systematic remapping on
Product-Matrix codes.

Traditionally, remapping is viewed as just decoding from the first
$k$ nodes, as discussed in Section~\ref{sec:sys-remap}.
In this case, understanding decoding is sufficient to understand systematic
remapping.
This is well-suited for classical codes, where the message-space and the code-space have
the same structure.
However, this is not true for Product-Matrix Codes.

In Product-Matrix Codes, encoding takes a (structured) message-matrix $M$ to a code-matrix $C$.
In the example code above, encoding the data of the first $k=4$ nodes is
a map:
$$
M =
\mqty[  m_0 & m_1 & m_2\\
        m_1 & m_3 & m_4\\
        m_2 & m_4 & m_5\\
    \hline
        m_6 & m_7 &  m_8\\
        m_7 & m_9 &  m_{10}\\
        m_8 & m_{10} & m_{11}]
\to
C =
\mqty[  c_0 & c_1 &  c_2\\
        c_3 & c_4 &  c_5\\
        c_6 & c_7 & c_8\\
        c_9 & c_{10} &  c_{11}]
$$
And decoding the $B=12$ message symbols from the first $k=4$ nodes
is the inverse map $C \to M$, whose explicit structure
follows from the decoding algorithm in \cite{PMC}.
However, understanding the explicit structure of the decoding map
does not immediately aid in understanding systematic-remapping.
This is because remapping is most naturally viewed as a transformation between
message-matrices $M \to \bM$.

We address the above challenge by presenting a framework for understanding
systematic remapping for product-matrix codes, and we further use
this to construct \pmmsr{} codes which \emph{remain sparse after
systematic remapping.}
An example of this construction is provided below.

\subsection{Sparse, Systematic \pmmsr{} Code}
In Sections~\ref{sec:explicit1} and~\ref{sec:explicit2}, we present explicit
constructions of sparse systematic \pmmsr{} codes.
Here we show the code construction presented
in Section~\ref{sec:explicit1}, instantiated for the same parameters
as the example of Section~\ref{sec:example_ex}: $[n=8, k=4, d=6]$.

The encoding matrix $\Psi'$ is chosen as:

$$
\Psi'=
\mqty[
1 & 0 & 0 & 1 & 0 & 0\\
0 & 1 & 0 & 0 & 8 & 0\\
0 & 0 & 1 & 0 & 0 & 5\\
4 & 5 & 4 & 3 & 1 & 3\\
4 & 2 & 10 & 5 & 8 & 7\\
3 & 10 & 9 & 10 & 4 & 8\\
4 & 4 & 2 & 8 & 8 & 4\\
10 & 3 & 1 & 5 & 7 & 6\\
]
$$

This yields the following (non-systematic) generator matrix:
$$
G'=
\smqty[
1 & 0 & 0 & 0 & 0 & 0 & 1 & 0 & 0 & 0 & 0 & 0\\
0 & 1 & 0 & 0 & 0 & 0 & 0 & 1 & 0 & 0 & 0 & 0\\
0 & 0 & 1 & 0 & 0 & 0 & 0 & 0 & 1 & 0 & 0 & 0\\
0 & 1 & 0 & 0 & 0 & 0 & 0 & 8 & 0 & 0 & 0 & 0\\
0 & 0 & 0 & 1 & 0 & 0 & 0 & 0 & 0 & 8 & 0 & 0\\
0 & 0 & 0 & 0 & 1 & 0 & 0 & 0 & 0 & 0 & 8 & 0\\
0 & 0 & 1 & 0 & 0 & 0 & 0 & 0 & 5 & 0 & 0 & 0\\
0 & 0 & 0 & 0 & 1 & 0 & 0 & 0 & 0 & 0 & 5 & 0\\
0 & 0 & 0 & 0 & 0 & 1 & 0 & 0 & 0 & 0 & 0 & 5\\
4 & 5 & 4 & 0 & 0 & 0 & 3 & 1 & 3 & 0 & 0 & 0\\
0 & 4 & 0 & 5 & 4 & 0 & 0 & 3 & 0 & 1 & 3 & 0\\
0 & 0 & 4 & 0 & 5 & 4 & 0 & 0 & 3 & 0 & 1 & 3\\
    \hline
4 & 2 & 10 & 0 & 0 & 0 & 5 & 8 & 7 & 0 & 0 & 0\\
0 & 4 & 0 & 2 & 10 & 0 & 0 & 5 & 0 & 8 & 7 & 0\\
0 & 0 & 4 & 0 & 2 & 10 & 0 & 0 & 5 & 0 & 8 & 7\\
3 & 10 & 9 & 0 & 0 & 0 & 10 & 4 & 8 & 0 & 0 & 0\\
0 & 3 & 0 & 10 & 9 & 0 & 0 & 10 & 0 & 4 & 8 & 0\\
0 & 0 & 3 & 0 & 10 & 9 & 0 & 0 & 10 & 0 & 4 & 8\\
4 & 4 & 2 & 0 & 0 & 0 & 8 & 8 & 4 & 0 & 0 & 0\\
0 & 4 & 0 & 4 & 2 & 0 & 0 & 8 & 0 & 8 & 4 & 0\\
0 & 0 & 4 & 0 & 4 & 2 & 0 & 0 & 8 & 0 & 8 & 4\\
10 & 3 & 1 & 0 & 0 & 0 & 5 & 7 & 6 & 0 & 0 & 0\\
0 & 10 & 0 & 3 & 1 & 0 & 0 & 5 & 0 & 7 & 6 & 0\\
0 & 0 & 10 & 0 & 3 & 1 & 0 & 0 & 5 & 0 & 7 & 6\\
]
\qquad\sim\qquad
\smqty[
* &   &   &   &   &   & * &   &   &   &   &  \\
  & * &   &   &   &   &   & * &   &   &   &  \\
  &   & * &   &   &   &   &   & * &   &   &  \\
  & * &   &   &   &   &   & * &   &   &   &  \\
  &   &   & * &   &   &   &   &   & * &   &  \\
  &   &   &   & * &   &   &   &   &   & * &  \\
  &   & * &   &   &   &   &   & * &   &   &  \\
  &   &   &   & * &   &   &   &   &   & * &  \\
  &   &   &   &   & * &   &   &   &   &   & *\\
* & * & * &   &   &   & * & * & * &   &   &  \\
  & * &   & * & * &   &   & * &   & * & * &  \\
  &   & * &   & * & * &   &   & * &   & * & *\\
    \hline
* & * & * &   &   &   & * & * & * &   &   &  \\
  & * &   & * & * &   &   & * &   & * & * &  \\
  &   & * &   & * & * &   &   & * &   & * & *\\
* & * & * &   &   &   & * & * & * &   &   &  \\
  & * &   & * & * &   &   & * &   & * & * &  \\
  &   & * &   & * & * &   &   & * &   & * & *\\
* & * & * &   &   &   & * & * & * &   &   &  \\
  & * &   & * & * &   &   & * &   & * & * &  \\
  &   & * &   & * & * &   &   & * &   & * & *\\
* & * & * &   &   &   & * & * & * &   &   &  \\
  & * &   & * & * &   &   & * &   & * & * &  \\
  &   & * &   & * & * &   &   & * &   & * & *\\
]
$$

After systematic-remapping, the final generator matrix is:
\beq
\label{eqn:newGsys}
G'_{sys}=
\smqty[
1 & 0 & 0 & 0 & 0 & 0 & 0 & 0 & 0 & 0 & 0 & 0\\
0 & 1 & 0 & 0 & 0 & 0 & 0 & 0 & 0 & 0 & 0 & 0\\
0 & 0 & 1 & 0 & 0 & 0 & 0 & 0 & 0 & 0 & 0 & 0\\
0 & 0 & 0 & 1 & 0 & 0 & 0 & 0 & 0 & 0 & 0 & 0\\
0 & 0 & 0 & 0 & 1 & 0 & 0 & 0 & 0 & 0 & 0 & 0\\
0 & 0 & 0 & 0 & 0 & 1 & 0 & 0 & 0 & 0 & 0 & 0\\
0 & 0 & 0 & 0 & 0 & 0 & 1 & 0 & 0 & 0 & 0 & 0\\
0 & 0 & 0 & 0 & 0 & 0 & 0 & 1 & 0 & 0 & 0 & 0\\
0 & 0 & 0 & 0 & 0 & 0 & 0 & 0 & 1 & 0 & 0 & 0\\
0 & 0 & 0 & 0 & 0 & 0 & 0 & 0 & 0 & 1 & 0 & 0\\
0 & 0 & 0 & 0 & 0 & 0 & 0 & 0 & 0 & 0 & 1 & 0\\
0 & 0 & 0 & 0 & 0 & 0 & 0 & 0 & 0 & 0 & 0 & 1\\
8 & 2 & 4 & 5 & 0 & 0 & 10 & 0 & 0 & 10 & 0 & 0\\
0 & 2 & 0 & 4 & 10 & 3 & 0 & 9 & 0 & 0 & 5 & 0\\
0 & 0 & 4 & 0 & 0 & 9 & 8 & 3 & 7 & 0 & 0 & 9\\
9 & 9 & 6 & 3 & 0 & 0 & 9 & 0 & 0 & 4 & 0 & 0\\
0 & 4 & 0 & 7 & 9 & 2 & 0 & 4 & 0 & 0 & 9 & 0\\
0 & 0 & 3 & 0 & 0 & 1 & 1 & 2 & 10 & 0 & 0 & 8\\
9 & 10 & 2 & 3 & 0 & 0 & 5 & 0 & 0 & 7 & 0 & 0\\
0 & 1 & 0 & 9 & 6 & 6 & 0 & 2 & 0 & 0 & 4 & 0\\
0 & 0 & 7 & 0 & 0 & 4 & 4 & 6 & 9 & 0 & 0 & 1\\
1 & 6 & 6 & 5 & 0 & 0 & 8 & 0 & 0 & 5 & 0 & 0\\
0 & 5 & 0 & 1 & 9 & 6 & 0 & 2 & 0 & 0 & 1 & 0\\
0 & 0 & 6 & 0 & 0 & 7 & 1 & 6 & 9 & 0 & 0 & 9\\
]
\qquad\sim\qquad
\smqty[
* &   &   &   &   &   &   &   &   &   &   &  \\
  & * &   &   &   &   &   &   &   &   &   &  \\
  &   & * &   &   &   &   &   &   &   &   &  \\
  &   &   & * &   &   &   &   &   &   &   &  \\
  &   &   &   & * &   &   &   &   &   &   &  \\
  &   &   &   &   & * &   &   &   &   &   &  \\
  &   &   &   &   &   & * &   &   &   &   &  \\
  &   &   &   &   &   &   & * &   &   &   &  \\
  &   &   &   &   &   &   &   & * &   &   &  \\
  &   &   &   &   &   &   &   &   & * &   &  \\
  &   &   &   &   &   &   &   &   &   & * &  \\
  &   &   &   &   &   &   &   &   &   &   & *\\
* & * & * & * &   &   & * &   &   & * &   &  \\
  & * &   & * & * & * &   & * &   &   & * &  \\
  &   & * &   &   & * & * & * & * &   &   & *\\
* & * & * & * &   &   & * &   &   & * &   &  \\
  & * &   & * & * & * &   & * &   &   & * &  \\
  &   & * &   &   & * & * & * & * &   &   & *\\
* & * & * & * &   &   & * &   &   & * &   &  \\
  & * &   & * & * & * &   & * &   &   & * &  \\
  &   & * &   &   & * & * & * & * &   &   & *\\
* & * & * & * &   &   & * &   &   & * &   &  \\
  & * &   & * & * & * &   & * &   &   & * &  \\
  &   & * &   &   & * & * & * & * &   &   & *\\
]
\eeq
Notice that in this case (compared to \eqref{eqn:Gsys}) the sparsity is not
lost in systematic-remapping:
Each row of $G'_{sys}$ is still $d=6$-sparse.


\section{First Step towards Sparsity in \pmmsr{} Codes}
\label{sec:step1}

In this section we analyze a simple family of encoding matrices $\Psi$
which, after systematic remapping, results in codes with partial sparsity.
The tools developed here will be useful
in subsequent sections.

Recall the structure of the encoding matrix for $d=(2k-2)$ \pmmsr{} codes
from \eqref{eqn:psi}:
$$
\Psi = \mqty[\Phi & \Lambda \Phi].
$$

Consider a $d=(2k-2)$ \pmmsr{} code in which the first row of $\Phi$
is $e_1 = \mqty[1 & 0  & \dots & 0]$.
\footnote{
Here we assume that such codes exist, and analyze their properties.
Explicit constructions of such codes are presented in Section~\ref{sec:exp_codes}.
}
Then the encoding matrix is of the form
\beq
\Psi = \mqty[e_1 & \lambda e_1\\
              \Phi' & \Lambda' \Phi'].
\eeq
We will now show that under such \pmmsr{} codes,
the first symbol stored in every node is $d$-sparse
after systematic remapping.

Let $\Psi_k$ denote the first $k$ rows of $\Psi$,
that is, the encoding submatrix for the first $k$ nodes.
Then the first $k$ nodes store
\beq
\Ck = \Psi_k M.
\eeq
Let $\fe: M \to \Ck$ denote the above encoding function for the first $k$ nodes.
We represent systematic remapping as a linear transformation $\fremap: M \to \bM$ between
the original matrix $M$ and the resultant message matrix after transformation
$\bM$.
After the remapping, the first $k$ nodes become systematic (see
Section~\ref{sec:sys-remap}).
That is, the transformation $\fremap$ is such that if we encode
the first $k$ nodes using
message matrix $\bM$, we recover the original symbols of $M$ in matrix $\Ck$.
Equivalently, for a systematic code, the entire encoding transform:
\beq
M \too[\fremap] \bM \too[\fe] \Ck
\eeq
must act as an inclusion map $M \incl \Ck$.
This inclusion map ``unwraps'' the symmetric matrices in $M$ into one matrix
$\Ck$ with distinct message symbols.

To understand the interaction between the \pmmsr{} code and systematic remapping,
we will define an explicit inclusion map
$\finc$, and decompose the remapping $\fremap$ into two stages.
We first represent the matrix $M$ as the matrix $\Ck$ using the inclusion map
$\finc$, and then ``decode'' $\Ck$ into $\bM$ using the decoding function
$\fe^{-1}$.
Note that $\fe$ is invertible since it is an MDS encoding,
wherein all message symbols can be decoded from any $k$ nodes.
The remapping transform thus becomes
\beq
\label{eqn:fremap_def}
\fremap = \fe^{-1} \circ \finc
\eeq
In other words,
\beq
\label{eqn:fremap_def2}
\fremap: M \incl[\finc] \Ck \too[\fe^{-1}] \bM
\eeq

Thus the entire encoding transformation for the first $k$ nodes becomes
\beq
M \incl[\finc] \Ck \too[\fe^{-1}] \bM \too[\fe] \Ck
\eeq

Notice that this makes the entire encoding transform $M \to \Ck$ an
inclusion map (equal to $\finc$, in fact), thus resulting
in a systematic code as desired.

\begin{remark}
Any choice of inclusion map
in~\eqref{eqn:fremap_def} will yield a systematic remapping.
However, as we will see, our particular choice of $\finc$ will be convenient for
proving sparsity results.
\end{remark}

At a high level, the key ideas behind our approach for showing sparsity are
as follows.
\begin{enumerate}
\item For our choices of $\Psi$ and $\finc$,
the systematic remapping $\fremap: M \to \bM$ is such that
the first column of $\bM$ depends only on the first column of $M$
(Lemma~\ref{lem:syscol1}).

\item
The first stored symbol in node $i$ is
the $i^{th}$ row of $\Psi$ times the first column of $\bM$.
This depends only on the first column of $\bM$,
and therefore (through $\fremap$) depends only on the first column of $M$.
\end{enumerate}

And Lemma~\ref{lem:syscol1} holds because:
\begin{enumerate}
\item The ``decoding'', $\fe^{-1}: \Ck \to \bM$
is such that the first column of $\bM$ depends only on the first row and first
column of $\Ck$ (Lemma~\ref{lem:invcol1}).

\item
Our inclusion map $\finc: M \incl \Ck$ will be such that the
symbols in the first row/column of $\Ck$ correspond exactly to the first column of
$M$.
\end{enumerate}

The sparsity pattern of systematic remapping (Lemma~\ref{lem:syscol1}) is visualized below:


\newcommand{\s}{{$\cdot$}}
\renewcommand{\k}{\cellcolor{black!25}}
$$
\left[
\begin{squarecells}{3}
        \k\s & \s & \s \nl
        \k\s & \s & \s \nl
        \k\s & \s & \s \nl
    \hline
        \k\s & \s  & \s \nl
        \k\s & \s  & \s \nl
        \k\s & \s  & \s \nl
\end{squarecells}
\right]
\mathlarger{\incl[\finc]}
\left[
\begin{squarecells}{3}
        \k\s & \k\s & \k\s \nl
        \k\s & \s & \s \nl
        \k\s & \s & \s \nl
\hline
        \k\s & \s & \s \nl
\end{squarecells}
\right]
\mathlarger{\too[\fe^{-1}]}
\left[
\begin{squarecells}{3}
        \k\s & \s & \s \nl
        \k\s & \s & \s \nl
        \k\s & \s & \s \nl
    \hline
        \k\s & \s  & \s \nl
        \k\s & \s  & \s \nl
        \k\s & \s  & \s \nl
\end{squarecells}
\right]
$$

$$\fremap: M \incl[\finc] \Ck \too[\fe^{-1}] \bM$$

We now consider each component of the systematic remapping transformation in
detail, and then prove the sparsity of the entire encoding.

\subsection{The Triangular Inclusion Map}
\label{sec:inclusion}
Here we will define the inclusion map $\finc$,
termed the ``triangular inclusion map.''
Recall from Section~\ref{sec:PMC}
that the message matrix in $d=(2k-2)$ \pmmsr{} codes
is of the form $M = \mqty[\Sa \\ \Sb]$, where $\Sa$ and $\Sb$ are symmetric
matrices of message symbols.

To map $M \incl \Ck$ by inclusion, place the upper-triangular half of $S^a$ on the upper-triangular
half of $\Ck$, including the diagonal. Then place the lower-triangular half of
$S^b$ on the lower-triangular half of $\Ck$, excluding the diagonal.
Finally, place the diagonal of $S^b$ on the last row of $\Ck$.
For example,
consider a \pmmsr{} code with $k=4, d=6$, for which $\alpha = 3$
and number of message-symbols $B=12$.
The triangular inclusion map $\finc$ in this case is:

\renewcommand{\r}{\cellcolor{red!25}}
\newcommand{\g}{\cellcolor{green!25}}
\renewcommand{\b}{\cellcolor{blue!25}}

\[
M = \left[
\begin{squarecells}{3}
        \r0 & \r1 & \r2 \nl
        1 & \r3 & \r4 \nl
        2 & 4 & \r5 \nl
\hline
        \g6 & 7 & 8 \nl
        \b7 & \g9 & 10 \nl
        \b8 & \b10 & \g11 \nl
\end{squarecells}
\right]
\mathlarger{\incl}
~\Ck =
\left[
\begin{squarecells}{3}
        \r0 & \r1 & \r2 \nl
        \b7 & \r3 & \r4 \nl
        \b8 & \b10 & \r5 \nl
        \g6 & \g9 & \g11 \nl
\end{squarecells}
\right]
\]

In the above matrices, numbers refer to symbol indices.
Notice that symbols in column $i$ ($0 \leq i \leq 2$) of $M$ correspond exactly to symbols
in row $i$ and column $i$ of $\Ck$.

\subsection{The Inverse Map}
Here we will consider the structure of the inverse map
$\fe^{-1}: \Ck \to M$, and show that it has a particular sparsity
pattern.

\begin{lemma}
\label{lem:invcol1}
In the inverse transform $\fe^{-1}: \Ck \to M$, the first column of $M$
depends only on the first row and first column of $\Ck$.
\end{lemma}

\proof
Let $\psi_1$ denote the first row of $\Psi$.
In the encoding transform $\fe: M \to \Ck$, notice that the first row and first column
of $\Ck$ depend only on the first column of $M$:
\begin{itemize}
    \item The first row of $\Ck$ is $\psi_1$ times $M$.
        Since $\psi_1 = \mqty[e_1 & \lambda e_1]$,
        this only involves symbols in the first row of $\Sa$
        and first row of $\Sb$. Or equivalently, the first column of $M$.
    \item The first column of $\Ck = \Psi M$ clearly depends only on the first
        column of $M$.
\end{itemize}
Therefore, we can consider the restriction of the map
$\fe: M \to \Ck$ to symbols in the first column of $M$, and the first
row/column of $\Ck$. There are $d$ symbols in both domain and
co-domain. Further, this map is full-rank by construction, since it is a restriction of
MDS encoding. Therefore this map is invertible, and the first
column of $M$ can be recovered from the first row/column of $\Ck$. \qed

\subsection{Sparsity}
Here we combine the above maps and show that the entire encoding transform has a
certain sparsity.
The following lemma serves as our main tool.

\begin{lemma}
\label{lem:syscol1}
In the systematic-remapping transform $\fremap: M \to \bM$,
a symbol in the first column of $\bM$ only depends on symbols
in the first column of $M$.
\end{lemma}

\proof
From \eqref{eqn:fremap_def2}, we write
the remapping transform as
$M \incl[\finc] \Ck \too[\fe^{-1}] \bM$.
From Lemma~\ref{lem:invcol1},
the first column of $\bar M$ depends only on the
first row/column of $\Ck$.
And by the triangular inclusion map $\finc$
as defined in Section~\ref{sec:inclusion},
the first row/column of $\Ck$
corresponds to the first column of $M$. \qed

We can then show partial sparsity of the entire encoding:

\begin{theorem}
    Consider a $d=(2k-2)$ \pmmsr{} code in which the first row of $\Phi$
is $e_1 = \mqty[1 & 0  & \dots & 0]$.
When this code is made systematic,
the first symbol stored in every
node is $d$-sparse.
\end{theorem}

\proof
The first symbol of each node is a row of $\Psi$
times the first column of $\bM$.
But the first column of $\bM$ depends only on the first column of $M$
(by Lemma~\ref{lem:syscol1}), so the first symbol of each node is $d$-sparse
w.r.t. symbols in $M$. Essentially, the sparsity occurs because the sparsity patterns of the
following two transformations, restricted to the first column of $M$, are aligned:
\beq
M \to \bM \to \Psi \bM.
\eeq
\qed

\begin{remark}
An analogous argument shows that if
one of the first $k$ rows of $\Phi$ is $e_i$, then the $i$-th
symbol of every node is $d$-sparse.
\end{remark}

\begin{remark}
It may seem that the above sparsity argument only works with
our particular inclusion map $\finc$, but in fact it applies to any systematic
remapping.
Notice that the systematic remapping function
is unique up to permutation of
the $k\alpha$ message symbols.
Therefore, if some final encoded symbol is a function of $d$ original message
symbols for a particular systematic-remapping function, it will remain a function of some $d$ (permuted) message
symbols in any other systematic-remapping.
\end{remark}

\section{Explicit Sparse, Systematic \pmmsr{} Codes for $d=(2k-2)$}
\label{sec:explicit1}

In this section, we first consider a particular design of encoding matrices
$\Psi$,
and prove that, after systematic remapping, they yield \pmmsr{} codes in which each
encoded symbol is $d$-sparse.
We then present explicit constructions of such matrices.
Our analysis builds on the techniques presented in the previous section.

In this section, for simplicity of notation, we will write the matrix $\Ck$ as simply $C$,
so $C_{i,j}$ denotes the $(i,j)^{th}$ entry of $C_k$.

\subsection{Design of the Encoding Matrix and Sparsity}
\label{sec:design}

Consider a $d=(2k-2)$ \pmmsr{} code in which the first $\alpha$ rows of $\Phi$ form an Identity
matrix. In this case, the encoding matrix for the first $k$
nodes is of the form:
\beq
\Psi_{k}
= \mqty[
I & \Lambda\\
r^T & \lambda r^T]
\eeq
where $r$ is an $\alpha$-length vector.
We will show under such an encoding matrix, after systematic remapping,
every encoded symbol is $d$-sparse.

From the properties of \pmmsr{} encoding matrices discussed in
Section~\ref{sec:PMC}, we have:
\begin{itemize}
    \item Property 1: The diagonal entries of $\Lambda$ together with $\lambda$
        are all distinct.
    \item Property 2:
        All sub-matrices of $\smqty[I \\ r^T]$ are full-rank.
        In particular, all entries of $r^T$ are nonzero.
\end{itemize}

The corresponding encoding transform for the first $k$ nodes, $\fe: M \to C$, is:
\begin{align}
C &= \Psi_k M\\
&=
\mqty[I & \Lambda \\ r^T & \lambda r^T]
\mqty[\Sa \\ \Sb]\\
&=
\mqty[\Sa + \Lambda \Sb\\
r^T\Sa + \lambda r^T \Sb]\\
&:=
\mqty[C_1 \\ C_2]
\end{align}

\subsubsection{The Inverse Map}
Here we describe how to recover the message matrices $\Sa$ and $\Sb$ from $\Ck$, thus specifying the
explicit structure of the inverse map $\fe^{-1}$.

First, all the non-diagonal entries of $\Sa, \Sb$ can be found by solving:
\beq
\label{eqn:nondiag}
\begin{cases}
    C_{i,j} = \Sa_{i,j} + \lambda_i \Sb_{i,j}\\
    C_{j,i} = \Sa_{i,j} + \lambda_j \Sb_{i,j}
\end{cases}
\eeq
(Since $\lambda_i \neq \lambda_j$ by Property 1).

For the diagonal entries, we first compute $\Sa r$ and $\Sb r$ as follows.\\
First define the following two vectors, which can be computed directly from $C$:
\begin{align}
    c_1 &:= C_1 r = \Sa r + \Lambda \Sb r \\
    c_2 &:= C_2^T = \Sa r + \lambda \Sb r
\end{align}
Then $\Sa r$ and $\Sb r$ can be computed from:
\begin{align}
    \label{eqn:Sar}
    \Sa r &= (\Lambda - \lambda I)^{-1}(\Lambda c_2 - \lambda c_1)\\
    \Sb r &= (\Lambda - \lambda I)^{-1}(c_1 - c_2)
\end{align}
where the diagonal matrix $(\Lambda - \lambda I)$ is invertible by Property 1.


Now we compute the $i$-th diagonal entry of $\Sa$ from $\Sa r$.
Let $\Sa_i$ denote row $i$ of $\Sa$.
After computing $\Sa r$ as above, we can extract $\Sa_i r =\sum_j \Sa_{i,j} r_j$.
Then $\Sa_{i,i}$ can be computed as:
\beq
\label{eqn:Saii}
\Sa_{i,i} = (\Sa_i r - \sum_{j \neq i}\Sa_{i,j} r_j) / r_i
\eeq
Notice that the non-diagonal elements $\Sa_{i, j\neq i}$ are known,
and $r_i \neq 0$ by Property 2.
The diagonal elements of $\Sb$ can be recovered similarly from $\Sb r$.

\subsubsection{Sparsity}
Using the structure of the inverse map described above, together with the triangular inclusion map
defined in Section~\ref{sec:inclusion}, we will show that the entire encoding
transform is $d$-sparse.

Analogous to Lemma~\ref{lem:syscol1}, we first show
that the systematic-remapping transform has a certain sparsity.

\begin{lemma}
\label{lem:syscols}
In the systematic-remapping transform $\fremap: M \to \bM$,
the symbol $\bM_{i,j}$ only depends on symbols
in column $j$ of $M$.
\end{lemma}

\proof
First notice that the sparsity pattern of $\fe^{-1}$, in recovering $\Sa$ from
$C$, is as follows:
\begin{itemize}
    \item Non-diagonal element $\Sa_{i,j}$ depends on elements $C_{i,j}$ and
        $C_{j,i}$, as in~\eqref{eqn:nondiag}.
    \item Diagonal element $\Sa_{i,i}$ depends on row $i$ and column $i$ of
        $C$.
        To see this, first compute all non-diagonal entries $\Sa_{i, j\neq i}$ from~\eqref{eqn:nondiag}
        using row $i$ and column $i$ of $C_1$. Then compute the $i$-th
        component of $\Sa r$ from \eqref{eqn:Sar}, using the $i$-th entry of
        $c_1$ and $c_2$.
        Finally, compute $\Sa_{i,i}$ from \eqref{eqn:Saii}.
\end{itemize}
And the same sparsity holds for recovering $\Sb$ from $C$ as well.

Now let $\bM = \mqty[\bar\Sa \\ \bar\Sb]$.
We will show that symbol $\bar\Sa_{i,j}$ depends only on
$\Sa_{i,j}$ and $\Sb_{i,j}$, and a symmetric argument holds for
$\bar\Sb_{i,j}$.
Writing the systematic-remapping as
$M \incl[\finc] C \too[\fe^{-1}] \bM$, there are two cases:

\begin{itemize}
\item
    Non-diagonal element $\bar{\Sa_{i,j}}$ depends on $C_{i,j}$ and $C_{j,i}$ which, by our inclusion map,
correspond to $\Sa_{i,j}$ and $\Sb_{i,j}$.

\item
    Diagonal element $\bar{\Sa_{j,j}}$ depend on row $j$ and column $j$ of $C$, which correspond to
column $j$ of $M$.
\end{itemize}
\qed

This allows us to show sparsity of the entire encoding.
\begin{theorem}
\label{thm:main-sparse}
Consider a $d=(2k-2)$ \pmmsr{} code in which the first $\alpha$ rows of $\Phi$ form an Identity
matrix.
When this code is made systematic,
each encoded symbol is $d$-sparse.
\end{theorem}
\proof
Each encoded symbol is a row of $\Psi$ times a column of $\bM$,
by the \pmmsr{} encoding of \eqref{eqn:CPsiM}.
But each column of $\bM$ depends only the corresponding column of $M$
(by Lemma~\ref{lem:syscols}).
Thus we conclude the final encoding $\Psi \bM$
is $d$-sparse w.r.t. symbols of $M$, since the two maps have aligned sparsity
patterns:
\beq
M \to \bar M \to \Psi \bar M.
\eeq
\qed

\subsection{Explicit Construction}
\label{sec:exp_codes}
We now present explicit constructions of matrices $\Psi$
which conform to the design of Section~\ref{sec:design}.
This yields explicit systematic $d=(2k-2)$ \pmmsr{} codes in which each encoded symbol is
$d$-sparse.

\begin{theorem}
\label{thm:sparseEnc}
Let
$\Psi = \mqty[\Phi & \Lambda \Phi]$
be the encoding matrix for a $d=(2k-2)$ \pmmsr{} code,
satisfying the properties mentioned in Section~\ref{sec:PMC}.
For example, we can let $\Psi$ be a Vandermonde matrix, as given in \cite{PMC}.
Let the $(\alpha \x \alpha)$ matrix $\Phi_\alpha$ denote the first $\alpha$ rows of $\Phi$.
Then the following encoding matrix:
\beq
\label{eqn:sparseEnc}
\Psi'
= \mqty[\Phi \Phi_\alpha^{-1} & \Lambda \Phi \Phi_\alpha^{-1}]
:= \mqty[\Phi' & \Lambda \Phi'].
\eeq
defines a $d=(2k-2)$ \pmmsr{} code in which, after systematic remapping,
each encoded symbol is $d$-sparse.
\end{theorem}
\proof
The matrix $\Psi'$ satisfies the properties of Section~\ref{sec:PMC},
since multiplication by full-rank $\Phi_\alpha^{-1}$ will not destroy the rank of any
submatrices of the original encoding matrix $\Psi$.
Therefore $\Psi'$ satisfies all properties of a \pmmsr{} encoding matrix.
Further, the first $\alpha$ rows of $\Phi'$ are the identity.
So by Theorem~\ref{thm:main-sparse}, we conclude that after systematic-remapping,
this code will remain $d$-sparse.\qed

In other words, if we represent the encoding procedure for this systematic code
as a $(n\alpha \x B)$ generator matrix $G$ mapping $B$ message symbols
to $n \alpha$ encoded symbols ($\alpha$ per node),
then each row of $G$ will be $d$-sparse.

\section{Sparsity in Systematic MSR Codes from Repair-By-Transfer}
\label{sec:general}

Sections~\ref{sec:step1} and~\ref{sec:explicit1} dealt with constructing sparse systematic \pmmsr{} codes.
In this section, we consider sparsity in more general systematic regenerating
codes.

\subsection{Background: MSR Codes and Repair-by-Transfer}
\label{sec:msr_background}
An $[n, k, d](\alpha, \beta)$
\emph{regenerating code} allows the message
to be stored across $n$ nodes, each storing $\alpha$ encoded symbols.
All the $B$ symbols can be recovered from the data stored in any $k$ of the
total $n$ nodes.
Further, any node's data may be exactly recovered by connecting to
any $d$ other nodes, and downloading $\beta \leq \alpha$ symbols from each.
The symbols transferred from a helper node during node repair
may in general be some arbitrary function of the data stored in it.

\emph{Minimum-Storage-Regenerating (MSR) codes} are regenerating codes
which are also MDS, and therefore satisfy
\beq
B = k \alpha.
\eeq
For example, an $[n, k, d]$ \pmmsr{} code is an
$[n, k, d](\alpha=d-k+1, \beta=1)$ MSR code.
The seminal work by Dimakis et. al. \cite{netcoding} shows that
for MSR codes, the parameters above must necessarily satisfy
\beq
\alpha = \beta(d-k+1).
\eeq

During a node-repair operation,
a helper node is said to perform \emph{repair-by-transfer (RBT)}
if it does not perform any computation and merely
transfers one of its $\alpha$ stored symbols to
the failed node.
We say a linear $[n,k,d](\alpha, \beta=1)$ MSR code supports
RBT with the \emph{RBT-SYS pattern}
if every node can help the first $\alpha$ nodes via RBT.

\subsection{Sparsity from Repair-by-Transfer}

We now present a general connection between sparsity and repair-by-transfer,
by showing that an MSR code with a certain RBT property must
necessarily be sparse.

Let $\C$ be a linear systematic MSR
$[n,k,d](\alpha,\beta=1)$ code of blocklength $B = k \alpha$,
with $(n\alpha \x B)$ generator matrix $G$.
\newcommand{\Gi}[1]{G^{(#1)}}
Let $\Gi{i}$ be the $(\alpha \x B)$ submatrix corresponding to the $i$-th node.

\begin{theorem}
\label{thm:rbt}
If $\C$ supports repair of a systematic node $\nu$ via RBT with helper nodes
comprising the remaining
$(k-1)$ systematic nodes and $d-(k-1) = \alpha$ other parity nodes,
then for each parity $i$, the corresponding generator-submatrix
$\Gi{i}$ has one row with sparsity $\leq d$.

In particular, the row of $\Gi{i}$ corresponding to the symbol transferred for the repair
of node $\nu$ is supported on at most the following coordinates.
\begin{itemize}
    \item The $\alpha$ coordinates corresponding to symbols
        stored by node $\nu$.
    \item For each of the other $(k-1)$ participating systematic nodes
        $\mu \neq \nu$:
        one coordinate corresponding to a symbol stored by node $\mu$.
\end{itemize}
\end{theorem}

\proof
Say systematic node 0 fails,
and is repaired via RBT by the $(k-1)$ other systematic nodes, and $\alpha$ other parity
nodes.
Each helper will send one of its $\alpha$ stored symbols.
For the systematic helpers, these symbols correspond
directly to message symbols -- let $S$ be the set of these message symbol indices.
Notice that $S$ is disjoint from the symbols that node 0 stores.
For the parity helpers, each transferred symbol is a linear combination of message
symbols. We claim that these linear combinations cannot be supported on more
than the message symbols that node 0 stores, and the set $S$.
That is, in total the support size can be at most $\alpha + (k-1) = d$.

Intuitively, Theorem~\ref{thm:rbt} holds because the symbols from systematic helpers can
only ``cancel interference'' in $(k-1)$ coordinates (of $S$), and the $\alpha$
parity helpers must allow the repair of node 0's $\alpha$ coordinates, and thus
cannot contain more interference. This concept of interference-alignment is made
precise in \cite{IA_in_RC}, and our Theorem~\ref{thm:rbt} follows as a corollary of
``Property 2 (Necessity of Interference Alignment)'' proved in Section VI.D of
\cite{IA_in_RC}. \qed

\begin{theorem}
    \label{thm:rbt2}
If $\C$ supports the \emph{RBT-SYS pattern},
then for each parity $i$, the corresponding generator-submatrix
$\Gi{i}$ has $\min(\alpha, k)$ rows that are $d$-sparse.
In particular, if $d \leq (2k-1)$, then all rows of $G$ are $d$-sparse.
\end{theorem}

\proof
In the \emph{RBT-SYS pattern}, each parity node $i$
helps the first $\alpha$ nodes via RBT, including $\min(\alpha, k)$ systematic nodes.
In each repair of a systematic node,
the row of $\Gi{i}$ corresponding to the RBT symbol sent is $d$-sparse (by Theorem 1).
This is true for each of the symbols sent to systematic nodes.
These transferred symbols correspond to distinct symbols stored in node $i$,
by Property 3, Section 6 of~\cite{IA_in_RC}, which states that these symbols
must be linearly independent.
Therefore, $\min(\alpha, k)$ rows of $\Gi{i}$ are $d$-sparse.

In particular, for an MSR code, $d \leq (2k-1)$ implies $\alpha \leq k$, so all
rows of $G$ are $d$-sparse in this regime.
\qed

\section{Explicit Sparse, Systematic \pmmsr{} Codes for $d > (2k-2)$}
\label{sec:explicit2}
Section~\ref{sec:general} provides a strong connection between
repair-by-transfer
and sparsity in systematic MSR codes. This connection allows us to construct
explicit sparse PM codes for $d > (2k-2)$.
First we review how to construct systematic \pmmsr{} codes which support the
\emph{RBT-SYS} pattern, from \cite{fast}.
We then review the notion of \emph{code shortening} for PM codes, from
\cite{PMC}.
We apply these tools with the results of Section~\ref{sec:general}
to present explicit systematic PM codes
in which all encoded symbols are $d$-sparse.

\subsection{Repair-By-Transfer (RBT) for \pmmsr{} Codes}
\label{sec:rbt}

Recall that in a code that supports the \emph{RBT-SYS} pattern,
if any of the first $\alpha$
nodes fail, every remaining node can help it by
simply transferring one of its stored
symbols.

For any $d \geq (2k-2)$, let $C = \Psi M$ be the code matrix of a \pmmsr{} code
$\C$.
Recall from Section~\ref{sec:background} that node $i$ stores a row $c_i^T = \psi_i^T M$.
To help repair node $f$, node $i$ sends $c_i^T \mu_f$, for some
repair vector $\mu_f$.
In helping the first $\alpha$ nodes, node $i$ would thus send
the $\alpha$ symbols
$c_i^T P$ where
\beq
\label{eqn:P}
P = \mqty[\mu_1 & \cdots & \mu_\alpha].
\eeq
Define the \emph{RBT-transformed code $\C'$}
as the code $\C$ where the data in each node is transformed
by $P$:
node $i$ now stores $c_i^T P$.
Hence the encoding procedure for $\C'$ results in the code matrix
\beq
C' = C P = \Psi M P.
\eeq
Notice that if $P$ is invertible, then $\C'$ shares the same
MDS and repair properties as $\C$.
Additionally, in $\C'$, node $i$ can help repair any of the first $\alpha$
nodes (say, node $j$) by simply transferring its $j\supth$ symbol:
$c_i^T \mu_j$.

For $d=(2k-2)$ \pmmsr{} codes, as reviewed in Section~\ref{sec:background},
the matrix $P = \Phi_\alpha^T$, which is invertible by construction.



\subsection{Code Shortening}
\label{sec:shorten}
The notion of \emph{code shortening} allows us to
construct $d > (2k-2)$ \pmmsr{} codes from a class of $d=(2k-2)$ \pmmsr{} codes.
Here we describe the \pmmsr{} code shortening of \cite{PMC}, stated in terms of
generator matrices.

For a generator matrix $G'$, consider the submatrix
$G$ obtained by omitting the first $t$ rows and
first $t$ columns of $G'$.
We refer to the code defined by $G$ as
\emph{the code $G'$, shortened by the first $t$ symbols}.



An $[n,k,d > 2k-2]$ \pmmsr{} code can be constructed
by simply shortening an $[n', k', d'=(2k'-2)]$ \pmmsr{} code, as follows.

\begin{lemma}
\label{lem:shorten}
(From Theorem 6 of \cite{PMC})
For any $[n, k, d > 2k-2]$, let $G'$ be the generator matrix of an
$[n' = n+i, k'= k+i, d'= d+i = (2k'-2)](\alpha, \beta)$ systematic \pmmsr{} code,
where $i := d-(2k-2)$.
Let $G$ be the submatrix of $G'$ obtained by omitting the first $i\alpha$ rows and
first $i\alpha$ columns.
Then $G$ defines a systematic $[n, k, d](\alpha, \beta)$ \pmmsr{} code.
\end{lemma}
\proof
Informally,
restricting to a submatrix as above can be thought of as considering
the subcode of $G'$ in which the first $i$ nodes store all $0$-symbols.
(Or equivalently, where the first $i\alpha$ message symbols are all $0$).
The regeneration and repair properties of $G'$ still hold in $G$
with $i$ less helpers ($k = k'-i, d = d'-i$) since the first $i$
``dummy nodes'' of $G'$ can be assumed to always send $0$ when participating in
regeneration or repair. Further, this new code still operates at the MSR point,
since the number of message symbols is $k'\alpha - i\alpha = k\alpha$.

Formally, the statement follows directly from Theorem 6 and Corollary 8 of \cite{PMC}.
\qed

\subsection{Explicit Construction}




Sparse systematic $d > (2k-2)$ MSR codes can be constructed by
RBT-transforming a $d'=(2k'-2)$ \pmmsr{} code, and then shortening appropriately.
The following theorem presents this result.

\begin{theorem}
\label{thm:supersparse}
Consider a $[n, k, d > (2k-2)]$ systematic \pmmsr{} code $\C$ constructed by
shortening a $[n'=n+i, k'=k+i, d' = (2k'-2)]$ systematic \pmmsr{} code $\C'$
that supports RBT-SYS, where $i:=(d-(2k-2))$.
Let $G$ denote the generator matrix for code $\C$.
Letting $\Gi{j}$ denote the $(\alpha \x k\alpha)$ submatrix of $G$ for node $j$,
the following sparsity holds for all nodes $j$.
\begin{itemize}
    \item The first $(d-2k+2)$ rows of $\Gi{j}$ are
        $k$-sparse.
    \item The remaining $(k-1)$ rows of $\Gi{j}$ are
        $d$-sparse.
\end{itemize}
\end{theorem}
\proof
By Lemma~\ref{lem:shorten}, the shortened generator matrix $G$
defines an $[n,k, d](\alpha, \beta)$ linear systematic MSR code.
The sparsity of $G$ follows from applying Theorem~\ref{thm:rbt}
to the code $G'$.
In particular, the first $i\alpha$ columns of $G'$
are omitted in $G$.
In the code $G'$, these columns correspond to symbols in the first $i$
systematic nodes -- we interchangeably denote these columns/nodes by set $N$.

Consider a row of $G'$ corresponding to a symbol transferred for the
repair (via RBT) of some systematic node $\nu \in N$.
By Theorem~\ref{thm:rbt},
the restriction of this row to columns outside $N$
must be $k$-sparse,
since it can only be supported on one symbol per systematic node $\mu \not\in N$.
There must be $|N| = i = (d-2k+2)$ such rows per $\Gi{j}$ since
the code $G'$ supports RBT-SYS, and symbols transferred from a given node
for the repair of two different nodes must be linearly independent
(in $d'=(2k'-2)$ \pmmsr{} codes)
by Property 3, Section 6 of~\cite{IA_in_RC}.

Now consider a row of $G'$ corresponding to a symbol transferred for the
repair (via RBT) of some systematic node $\nu \not\in N$.
By Theorem~\ref{thm:rbt},
the restriction of this row to columns outside $N$
must be $(\alpha + k-1)=d$-sparse,
since it can only be supported on the $\alpha$ symbols of $\nu$
plus one symbol per remaining systematic node $\mu \not\in N, \mu \neq \nu$.
This comprises the remaining rows of each $\Gi{j}$,
similarly by the RBT-SYS property and Property 3, Section 6 of~\cite{IA_in_RC}.
\qed

\begin{remark}
It is interesting to note that the sparsity provided by
the codes of Theorem~\ref{thm:supersparse} is greater than
the sparsity guaranteed by a generic $[n, k, d > (2k-2)](\alpha, \beta=1)$
linear systematic MSR code that supports RBT-SYS.
By Theorem~\ref{thm:rbt2}, such a code would be such that the first $k$
symbols stored in every node are $d$-sparse, while the remaining symbols may be
dense.
\footnote{
It turns out that the unified \pmmsr{} codes presented in
\cite{unified} also have a certain degree of inherent sparsity,
although not as sparse as the codes of Theorem~\ref{thm:supersparse}.
It can be shown using an inclusion map argument
that the codes of \cite{unified}, in systematic form, have the following sparsity pattern:
\emph{the last $(\alpha - k)$ symbols stored in every node are $k$-sparse.}
Interestingly, the RBT-transformed version of these codes
have essentially the complementary sparsity pattern (by the present remark).
}
\end{remark}


\section{Equivalence in Sparse Systematic \pmmsr{} Code Constructions}
\label{sec:equivalence}

The previous sections present
two different ways of a constructing sparse $d=(2k-2)$ \pmmsr{} code
from a given $d=(2k-2)$ \pmmsr{} code:
\begin{enumerate}
    \item Apply the RBT-transformation of Section~\ref{sec:rbt}
        to yield a code that is sparse (by Theorem~\ref{thm:rbt2}).
    \item Transform the encoding matrix $\Phi$ to contain an identity block,
        as in Equation~\eqref{eqn:sparseEnc} of Theorem~\ref{thm:sparseEnc}.
\end{enumerate}
Interestingly, it turns out that these two constructions are equivalent up to a transform termed \emph{symbol-remapping}, which is defined below.

\emph{Symbol-remapping} is defined as any invertible transformation on the
message-space of a code. For example, systematic-remapping is a special case of
symbol-remapping for achieving systematic codes.
Two codes with encoding functions $f_1$ and $f_2$ are
\emph{equivalent up to symbol-remapping} if
\beq
f_1 = f_2 \circ T
\eeq
for some invertible transform $T$.

\begin{theorem}
For a given $d=(2k-2)$ \pmmsr{} code $\C$
with encoding matrix
$\Psi = \begin{bmatrix}\Phi & \Lambda\Phi \end{bmatrix}$,
consider a related code $\C'$ wherein the data in each node is
further transformed by an invertible linear transformation $P$.
That is, the entire encoding operation is $C' = \Psi M P$.
Then $\C'$ is equivalent to a \pmmsr{} code with the
below encoding matrix $\Psi'$ up to symbol-remapping.
\beq
\Psi' := \begin{bmatrix}\Phi P^{-T} & \Lambda\Phi P^{-T} \end{bmatrix}
\eeq
\end{theorem}

\proof
\renewcommand{\t}{\widetilde}
Consider transforming each message-submatrix $\Sa$ and $\Sb$ by
\beq
\Sa \to \t\Sa := P^{-T} \Sa P^{-1}
\eeq
Notice that this transformation is invertible and
preserves symmetry, so it is a symbol-remapping
on the message-space of \pmmsr{} codes.

If we then encode $\C'$ using message matrices $\t\Sa$ and $\t\Sb$,
the entire encoding operation will be:
\begin{align}
\Psi \begin{bmatrix} \t\Sa \\ \t\Sb \end{bmatrix} P
&= \Psi \begin{bmatrix} \t\Sa P \\ \t\Sb P \end{bmatrix}\\
&= \Psi \begin{bmatrix} P^{-T}\Sa \\ P^{-T}\Sb \end{bmatrix}\\
&=\begin{bmatrix}\Phi P^{-T} & \Lambda\Phi P^{-T} \end{bmatrix}
\begin{bmatrix} \Sa \\ \Sb \end{bmatrix}\\
&=\Psi'\begin{bmatrix} \Sa \\ \Sb \end{bmatrix}
\end{align}

The above form is native \pmmsr{} encoding with the original message
matrix
$M = \begin{bmatrix} \Sa \\ \Sb \end{bmatrix}$,
and the new encoding matrix
$\Psi' := \begin{bmatrix}\Phi P^{-T} & \Lambda\Phi P^{-T} \end{bmatrix}.$
\qed

Notice that if $P$ is chosen to support RBT-SYS (as in Section~\ref{sec:rbt}), then $P^T$ will be the first
$\alpha$ rows of $\Phi$, and the encoding matrix
\beq
\Psi'
= \mqty[\Phi P^{-T} & \Lambda\Phi P^{-T}]
= \mqty[\Phi \Phi_\alpha^{-1} & \Lambda\Phi \Phi_\alpha^{-1}]
\eeq
is identical to the encoding matrix \eqref{eqn:sparseEnc} of the explicit sparse codes
of Theorem~\ref{thm:sparseEnc}.

Thus these two methods of constructing sparse codes are equivalent up to
symbol-remapping.

\bibliography{bib}{}

\begin{thebibliography}{10}
\providecommand{\url}[1]{#1}
\csname url@samestyle\endcsname
\providecommand{\newblock}{\relax}
\providecommand{\bibinfo}[2]{#2}
\providecommand{\BIBentrySTDinterwordspacing}{\spaceskip=0pt\relax}
\providecommand{\BIBentryALTinterwordstretchfactor}{4}
\providecommand{\BIBentryALTinterwordspacing}{\spaceskip=\fontdimen2\font plus
\BIBentryALTinterwordstretchfactor\fontdimen3\font minus
  \fontdimen4\font\relax}
\providecommand{\BIBforeignlanguage}[2]{{%
\expandafter\ifx\csname l@#1\endcsname\relax
\typeout{** WARNING: IEEEtran.bst: No hyphenation pattern has been}%
\typeout{** loaded for the language `#1'. Using the pattern for}%
\typeout{** the default language instead.}%
\else
\language=\csname l@#1\endcsname
\fi
#2}}
\providecommand{\BIBdecl}{\relax}
\BIBdecl

\bibitem{gfs}
S.~Ghemawat, H.~Gobioff, and S.-T. Leung, ``The google file system,'' in
  \emph{ACM SIGOPS operating systems review}, vol.~37, no.~5.\hskip 1em plus
  0.5em minus 0.4em\relax ACM, 2003, pp. 29--43.

\bibitem{hdfs-raid1}
B.~Fan, W.~Tantisiriroj, L.~Xiao, and G.~Gibson, ``Diskreduce: Raid for
  data-intensive scalable computing,'' in \emph{Proceedings of the 4th Annual
  Workshop on Petascale Data Storage}.\hskip 1em plus 0.5em minus 0.4em\relax
  ACM, 2009, pp. 6--10.

\bibitem{hdfs-raid2}
D.~Borthakur, R.~Schmidt, R.~Vadali, S.~Chen, and P.~Kling, ``Hdfs raid,'' in
  \emph{Hadoop User Group Meeting}, 2010.

\bibitem{rashmi2013hotstorage}
K.~V. Rashmi, N.~B. Shah, D.~Gu, H.~Kuang, D.~Borthakur, and K.~Ramchandran,
  ``A solution to the network challenges of data recovery in erasure-coded
  distributed storage systems: A study on the {F}acebook warehouse cluster,''
  in \emph{Proc. USENIX HotStorage}, Jun. 2013.

\bibitem{vulimiri2012more}
A.~Vulimiri, O.~Michel, P.~Godfrey, and S.~Shenker, ``More is less: Reducing
  latency via redundancy,'' in \emph{11th ACM Workshop on Hot Topics in
  Networks}, Oct. 2012, pp. 13--18.

\bibitem{liang2014fast}
G.~Liang and U.~C. Kozat, ``Fast cloud: Pushing the envelope on delay
  performance of cloud storage with coding,'' \emph{Networking, IEEE/ACM
  Transactions on}, vol.~22, no.~6, pp. 2012--2025, 2014.

\bibitem{ananthanarayanan2012let}
G.~Ananthanarayanan, A.~Ghodsi, S.~Shenker, and I.~Stoica, ``Why let resources
  idle? {A}ggressive cloning of jobs with {D}olly,'' in \emph{USENIX HotCloud},
  Jun. 2012.

\bibitem{dean2013tail}
J.~Dean and L.~A. Barroso, ``The tail at scale,'' \emph{Communications of the
  ACM}, vol.~56, no.~2, pp. 74--80, 2013.

\bibitem{shah2013redundant}
N.~B. Shah, K.~Lee, and K.~Ramchandran, ``When do redundant requests reduce
  latency?'' in \emph{Communication, Control, and Computing (Allerton), 2013
  51st Annual Allerton Conference on}.\hskip 1em plus 0.5em minus 0.4em\relax
  IEEE, 2013, pp. 731--738.

\bibitem{lin-costello}
S.~Lin and D.~Costello, \emph{Error Control Coding: Fundamentals and
  Applications}, ser. Prentice-Hall computer applications in electrical
  engineering series.\hskip 1em plus 0.5em minus 0.4em\relax Prentice-Hall,
  1983.

\bibitem{MBR_pt}
K.~V. Rashmi, N.~B. Shah, P.~V. Kumar, and K.~Ramchandran, ``Explicit and
  optimal exact-regenerating codes for the minimum-bandwidth point in
  distributed storage,'' in \emph{Proc. IEEE International Symposium on
  Information Theory (ISIT)}, Austin, Jun. 2010, pp. 1938--1942.

\bibitem{MISERjournal}
N.~B. Shah, K.~V. Rashmi, P.~V. Kumar, and K.~Ramchandran, ``Interference
  alignment in regenerating codes for distributed storage: Necessity and code
  constructions,'' \emph{IEEE Transactions on Information Theory}, vol.~58,
  no.~4, pp. 2134--2158, Apr. 2012.

\bibitem{tamo2013zigzag}
I.~Tamo, Z.~Wang, and J.~Bruck, ``Zigzag codes: {MDS} array codes with optimal
  rebuilding,'' \emph{Information Theory, IEEE Transactions on}, vol.~59,
  no.~3, pp. 1597--1616, 2013.

\bibitem{papailiopoulos2013repairTransactions}
D.~Papailiopoulos, A.~Dimakis, and V.~Cadambe, ``Repair optimal erasure codes
  through {H}adamard designs,'' \emph{IEEE Transactions on Information Theory},
  vol.~59, no.~5, pp. 3021--3037, May 2013.

\bibitem{cadambe2011polynomial}
V.~Cadambe, C.~Huang, J.~Li, and S.~Mehrotra, ``Polynomial length {MDS} codes
  with optimal repair in distributed storage,'' in \emph{Forty Fifth Asilomar
  Conference on Signals, Systems and Computers}, Nov. 2011, pp. 1850--1854.

\bibitem{PMC}
K.~Rashmi, N.~B. Shah, and P.~V. Kumar, ``Optimal exact-regenerating codes for
  distributed storage at the msr and mbr points via a product-matrix
  construction,'' \emph{Information Theory, IEEE Transactions on}, vol.~57,
  no.~8, pp. 5227--5239, 2011.

\bibitem{netcoding}
A.~G. Dimakis, P.~Godfrey, Y.~Wu, M.~J. Wainwright, and K.~Ramchandran,
  ``Network coding for distributed storage systems,'' \emph{Information Theory,
  IEEE Transactions on}, vol.~56, no.~9, pp. 4539--4551, 2010.

\bibitem{jiekak2012system}
S.~Jiekak, A.-M. Kermarrec, N.~Le~Scouarnec, G.~Straub, and A.~Van~Kempen,
  ``Regenerating codes: A system perspective,'' \emph{ACM SIGOPS Operating
  Systems Review}, vol.~47, no.~2, pp. 23--32, 2013.

\bibitem{fast-pm-codes}
N.~L. Scouarnec, ``Fast product-matrix regenerating codes,'' \emph{CoRR}, vol.
  abs/1412.3022, 2014.

\bibitem{allerton09}
K.~Rashmi, N.~B. Shah, P.~V. Kumar, and K.~Ramchandran, ``Explicit construction
  of optimal exact regenerating codes for distributed storage,'' in
  \emph{Communication, Control, and Computing, 2009. Allerton 2009. 47th Annual
  Allerton Conference on}.\hskip 1em plus 0.5em minus 0.4em\relax IEEE, 2009,
  pp. 1243--1249.

\bibitem{fast}
K.~Rashmi, P.~Nakkiran, J.~Wang, N.~B. Shah, and K.~Ramchandran, ``Having your
  cake and eating it too: Jointly optimal erasure codes for i/o, storage, and
  network-bandwidth,'' in \emph{13th USENIX Conference on File and Storage
  Technologies (FAST 15)}, Santa Clara, CA, 2015.

\bibitem{rouayheb2010fractional}
S.~{El Rouayheb} and K.~Ramchandran, ``Fractional repetition codes for repair
  in distributed storage systems,'' in \emph{{Allerton Conference on Control,
  Computing, and Communication}}, Urbana-Champaign, Sep. 2010.

\bibitem{pawardress}
S.~Pawar, N.~Noorshams, S.~El~Rouayheb, and K.~Ramchandran, ``Dress codes for
  the storage cloud: Simple randomized constructions,'' in \emph{Proc. IEEE
  International Symposium on Information Theory (ISIT)}, St. Petersburg, Aug.
  2011.

\bibitem{hu2013analysis}
Y.~Hu, P.~P. Lee, and K.~W. Shum, ``Analysis and construction of functional
  regenerating codes with uncoded repair for distributed storage systems,'' in
  \emph{INFOCOM, 2013 Proceedings IEEE}.\hskip 1em plus 0.5em minus 0.4em\relax
  IEEE, 2013, pp. 2355--2363.

\bibitem{rashmi2014hitchhiker}
K.~Rashmi, N.~B. Shah, D.~Gu, H.~Kuang, D.~Borthakur, and K.~Ramchandran, ``A
  hitchhiker's guide to fast and efficient data reconstruction in erasure-coded
  data centers,'' in \emph{Proceedings of the 2014 ACM conference on
  SIGCOMM}.\hskip 1em plus 0.5em minus 0.4em\relax ACM, 2014, pp. 331--342.

\bibitem{unified}
S.-J. Lin, W.-H. Chung, Y.~S. Han, and T.~Y. Al-Naffouri, ``A unified form of
  exact-msr codes via product-matrix frameworks,'' \emph{Information Theory,
  IEEE Transactions on}, vol.~61, no.~2, pp. 873--886, 2015.

\bibitem{rskgeneral}
M.~Kurihara and H.~Kuwakado, ``{Generalization of Rashmi-Shah-Kumar
  Minimum-Storage-Regenerating Codes},'' \emph{arXiv preprint arXiv:1309.6701},
  2013.

\bibitem{IA_in_RC}
N.~B. Shah, K.~Rashmi, P.~V. Kumar, and K.~Ramchandran, ``Interference
  alignment in regenerating codes for distributed storage: Necessity and code
  constructions,'' \emph{Information Theory, IEEE Transactions on}, vol.~58,
  no.~4, pp. 2134--2158, 2012.

\end{thebibliography}
\bibliographystyle{IEEEtran}

\end{document}